\documentclass[conference]{IEEEtran}

\usepackage{amsmath,amssymb}
\usepackage{iftex}

\usepackage{cite}
\usepackage{graphicx}
\usepackage{float}
\usepackage{booktabs}
\usepackage{xurl}
\usepackage{xcolor}
\usepackage{tikz}
\usetikzlibrary{arrows.meta,positioning,fit,calc}

\definecolor{HCNavy}{HTML}{17365D}
\definecolor{HCTeal}{HTML}{0B7A75}
\definecolor{HCOrange}{HTML}{D97706}
\definecolor{HCLightBlue}{HTML}{EAF2F8}
\definecolor{HCLightTeal}{HTML}{E8F5F3}
\definecolor{HCLightOrange}{HTML}{FFF3E0}
\definecolor{HCGray}{HTML}{F2F4F6}

\begin{document}
\raggedbottom
\bstctlcite{IEEEFullAuthorNames}

\title{N-1 Security constrained Hosting Capacity Assessment for Ultra Large-scale Data Center with Workload Flexibility }

\author{\IEEEauthorblockN{Jiakai Wu, Pengyu Ren, Yifan Wang, and Wei Sun}
\IEEEauthorblockA{School of Engineering, The University of Edinburgh\\
Edinburgh EH9 3FB, Scotland, United Kingdom}}

\maketitle

\begin{abstract}
Transmission network security can substantially alter where large-scale data centers can connect and how much workload flexibility can unlock. This paper develops a hosting-capacity framework that couples computing demand, batch workload scheduling, and N-1 security constraints. A modified IEEE 118-bus case study reveals that two data center sites with identical hosting capacities under normal operation differ by 874.5 MW under the contingencies operation when workload flexibility is enabled. Six-hour batch deferral increases N-1 hosting capacity by up to 24.5\% relative to immediate execution. These gains arise from reducing demand at critical hours while accommodating deferred tasks within subsequent transmission limits and deadlines. At bus 92, a four-hour window captures 90.9\% of the six-hour capacity gain. The findings establish workload flexibility as a location-dependent connection resource whose capacity value requires joint assessment of contingency exposure and deadline constrainted workload recovery, beyond the power reduction available at a single critical hour.
\end{abstract}


\begin{figure*}[!t]
    \centering
    \includegraphics[width=.72\textwidth]{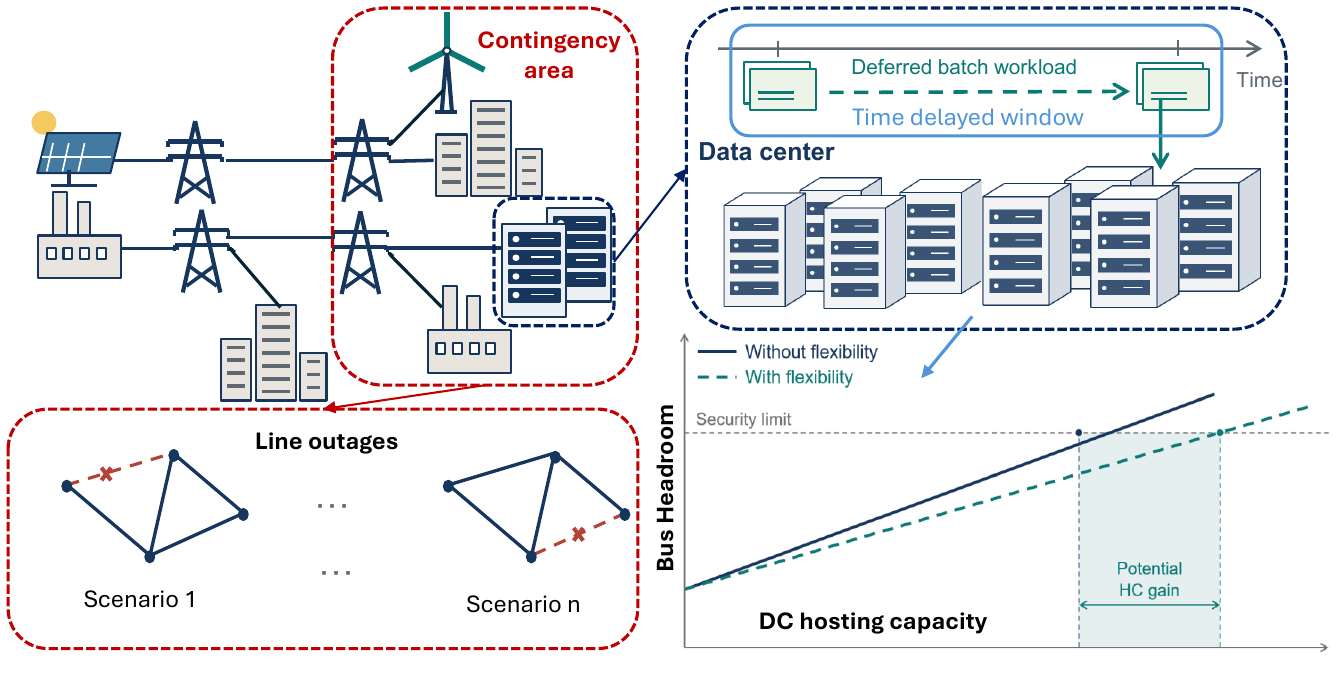}
    \caption{Illustration of data center hosting capacity under line
    outages with workload flexibility.}
    \label{fig:batch_security}
\end{figure*}

\section{Introduction}

Artificial intelligence (AI) is accelerating the growth and concentration
of data center electricity demand. The International Energy Agency (IEA)
estimated global consumption at 415~TWh in 2024 and projected approximately
945~TWh by 2030 in its 2025 outlook \cite{IEA2025EnergyAI}.
Its October 2026 report, \emph{Powering Artificial Intelligence},
identifies network capacity and connection queues as barriers to
integration and discusses compute flexibility as a means of facilitating
grid connections \cite{IEA2026PoweringAI}.
For transmission-connected facilities, feasible connection capacity
depends on both the location and timing of demand and the network's
ability to supply it following credible outages. Gird planning
must therefore consider network security alongside the operating
flexibility of the connected facility .

Workload flexibility allows data centers to adjust electricity demand
while meeting computing service requirements. Latency-sensitive tasks
require prompt execution, whereas eligible batch tasks can be deferred
within completion deadlines. Such scheduling can accommodate additional
capacity by shifting demand away from constrained hours. Under preventive
N$-$1 operation, however, the scheduled demand must also remain feasible
following each selected single-line outage, including during the hours
when deferred work is executed \cite{Giraud2024SCOPF}.
The value of workload flexibility therefore requires a joint
assessment of computing service obligations, batch deadlines, and
post-contingency transmission limits.

Hosting capacity (HC) quantifies the demand a network can accommodate
while satisfying specified operating limits
\cite{Benzerga2025HostingCapacity}. For transmission-connected data
centers, Manassero et al. \cite{Manassero2026ResilienceAware} develop
a resilience-aware allocation methodology that considers transmission
HC and contingency scenarios to support connection planning.
Kim et al. \cite{Kim2026FlexibilityAware} propose a planner-initiated
siting framework that combines reliability screening, market-impact
assessment, and location ranking under different operating flexibility
options. These studies connect data center integration with network
capacity, location, and operating flexibility. Transmission security has also been addressed through
security-constrained optimal power flow (SCOPF).
Giraud et al. \cite{Giraud2024SCOPF} combine line outage distribution
factors with constraint-driven learning to determine preventive
dispatch under multiple line outages. Assessing the capacity benefit of data center workload flexibility requires an explicit representation of data center and workload model.

Research on workload flexibility provides models for linking computing
service to electrical demand. Wang et al. \cite{Wang2026DemandFlexibility} develop a
chance-constrained bi-level model that couples power capping
with job scheduling to quantify demand flexibility subject
to service-level requirements.
Ren et al. \cite{Ren2026FrequencySupport} quantify the frequency-support
potential of data center flexibility by incorporating fast frequency
response into frequency-secured unit commitment.
Tsiligkaridis et al. \cite{Tsiligkaridis2025DistributedDispatch}
integrate workload shifting and server utilization into economic
dispatch with transmission constraints and queueing-based
service-quality costs. Li et al.
\cite{Li2024ComputationPower} develop a computation-power
coupling model and workload scheduling mechanism for collaborative
optimization between geographically distributed data centers and
distribution networks. Existing studies have investigated flexibility modelling and coordinated operation, but have not fully established how workload flexibility increases data centre hosting capacity and how this flexibility value varies with transmission security requirements and batch workload deferral windows.

To address these questions, this paper makes the following contributions.
\begin{enumerate}
\item An optimization framework is developed to assess data center hosting capacity under preventive N-1 transmission constraints. The framework jointly determines connection capacity and hourly operation by linking workload arrivals, deadline-constrained batch execution, server capability, and electrical demand.
\item The impact of N-1 transmission security constraints on the capacity value of workload flexibility is characterized. The analysis relates changes in workload flexibility value to the different capacity reductions imposed by contingency constraints between flexible and inflexible workload operation.
\item The operational mechanism and deadline dependence of security constrained hosting capacity gains are demonstrated. It analysis investigates how workload scheduling accommodates larger data center capacities while satisfying the service quality and quantifies the capacity benefit of different completion deadlines.
\end{enumerate}

\section{Modelling}
\label{sec:methodology}
To achieve electricity-computation co-optimization, the problem is formulated from a coordinated system planning perspective. 
Fig.~\ref{fig:batch_security}
This section formulates the maximum data center hosting capacity at the transmission network under N-1 security constraints. As illustrated in Fig. 1, a line outage redistributes power flows across the remaining network, potentially restricting the capacity available to accommodate the data center. Increasing the data center capacity increases both computing tasks and the workload to be served, while batch deferral reshapes the power demand within completion deadlines. The network must supply this demand while respecting line limits in the normal operation and contingency situation. 
\subsection{Data Center and Workload Flexibility Model}
Let \(\bar a_t\) be the workload arrival rate per MW and
\(\alpha_t^{\mathrm B}\) the batch share. Total workload and batch workload arrival rates are
\begin{equation}
W_t^0=H\bar a_t,\qquad W_t^{\mathrm{B}0}=\alpha_t^{\mathrm B}W_t^0.
\label{eq:capacity_linked_arrivals}
\end{equation}
Batch workloads comprise delay-tolerant computational tasks, such as
eligible AI training and data management jobs, whose execution can be
scheduled within prescribed completion windows. Unlike tasks requiring
immediate processing, these workloads may be deferred at the data
center, allowing their electrical demand to be redistributed across hours.
All arriving work must still be completed within its deadline, so deferral
changes the timing of computing demand while preserving the associated
service obligation.

Non-flexible tasks should be executed upon arrival. The maximum batch delay \(K\) is a prescribed integer constant fixed before each optimisation, with
\(1\leq K<T/2\). All workload rates are expressed in workload units per
hour. Let \(q_{t,\tau}\geq0\) denote workload arriving in hour \(t\) and
deferred to hour \(\tau\), where \(1\leq t<\tau\leq T\) and
\(\tau-t\leq K\). The executed batch workload \(w_t^{\mathrm B}\) includes
current arrivals retained for immediate processing and deferred work
recovered from earlier hours. Its balance is specified separately for the
initial, intermediate, and terminal intervals to account for the available
scheduling window at the horizon boundaries.

For the initial hours \(1\leq t\leq K\),
\begin{gather}
w_t^{\mathrm B}=W_t^{\mathrm{B}0}
+\sum_{\tau=1}^{t-1}q_{\tau,t}
-\sum_{\tau=t+1}^{t+K}q_{t,\tau},
\label{eq:batch_initial_balance}\\
\sum_{\tau=t+1}^{t+K}q_{t,\tau}\leq W_t^{\mathrm{B}0}.
\label{eq:batch_initial_limit}
\end{gather}
For the intermediate hours \(K+1\leq t\leq T-K\),
\begin{gather}
w_t^{\mathrm B}=W_t^{\mathrm{B}0}
+\sum_{\tau=t-K}^{t-1}q_{\tau,t}
-\sum_{\tau=t+1}^{t+K}q_{t,\tau},
\label{eq:batch_interior_balance}\\
\sum_{\tau=t+1}^{t+K}q_{t,\tau}\leq W_t^{\mathrm{B}0}.
\label{eq:batch_interior_limit}
\end{gather}
For the terminal hours \(T-K+1\leq t\leq T\),
\begin{gather}
w_t^{\mathrm B}=W_t^{\mathrm{B}0}
+\sum_{\tau=t-K}^{t-1}q_{\tau,t}
-\sum_{\tau=t+1}^{T}q_{t,\tau},
\label{eq:batch_terminal_balance}\displaybreak[1]\\
\sum_{\tau=t+1}^{T}q_{t,\tau}\leq W_t^{\mathrm{B}0}.
\label{eq:batch_terminal_limit}
\end{gather}
The incoming and outgoing sums represent recovery and deferral,
respectively. These balances impose zero initial backlog and completion by the end of the horizon, giving
\begin{gather}
\sum_{t=1}^{T}w_t^{\mathrm B}=\sum_{t=1}^{T}W_t^{\mathrm{B}0},
\label{eq:batch_completion}\\
w_t=W_t^0-W_t^{\mathrm{B}0}+w_t^{\mathrm B}.
\label{eq:processed_workload}
\end{gather}
Without workload flexibility, all \(q_{t,\tau}\) are fixed to zero and
\(w_t=W_t^0\). With flexibility, they are scheduling decisions within the
prescribed delay window.

Let \(m_t\) be the active-server count, \(\mu\) the processing rate per
fully utilised server, \(\bar u\) the utilisation limit, and \(\rho\) the
installed servers per MW. Processing and facility power obey
\begin{gather}
0\leq w_t\leq\mu\bar u m_t,\qquad 0\leq m_t\leq\rho H,
\label{eq:server_limits}\\
p_t^{\mathrm{IT}}=p^{\mathrm{idle}}m_t+
\frac{p^{\mathrm{peak}}-p^{\mathrm{idle}}}{\mu}w_t,
\label{eq:it_power_model}\\
p_t^{\mathrm{DC}}=\eta p_t^{\mathrm{IT}},\qquad
0\leq p_t^{\mathrm{DC}}\leq H.
\label{eq:facility_power_model}
\end{gather}
Here \(p^{\mathrm{idle}}\) and \(p^{\mathrm{peak}}\) are per-server powers
in MW \cite{Fan2007PowerProvisioning}, and \(\eta\) is the facility PUE
\cite{Shehabi2024DataCenterEnergy}. We set
\(\rho^{-1}=\eta[p^{\mathrm{idle}}+\bar u(p^{\mathrm{peak}}-p^{\mathrm{idle}})]\)
so the installed fleet draws \(H\) MW at its utilisation limit.
The minimum power for processing rate \(w\), attained at
\(m=w/(\mu\bar u)\), is
\begin{equation}
\underline p^{\mathrm{DC}}(w)=
\frac{\eta}{\mu}\left(
\frac{p^{\mathrm{idle}}}{\bar u}
+p^{\mathrm{peak}}-p^{\mathrm{idle}}\right)w.
\label{eq:minimum_import}
\end{equation}
Evaluating \eqref{eq:minimum_import} at \(w=W_t^0\) gives the
immediate-execution benchmark at the same facility capacity.

\subsection{Network Operation and N$-$1 Security Constraint}
Let \(\mathcal N\) and \(\mathcal L\) denote buses and branches, and
\(\mathcal G_i\subseteq\mathcal G\) and \(\mathcal R_i\subseteq\mathcal R\) the conventional and profile-based
sources at bus \(i\). With native demand \(d_{i,t}\), nodal balance is
\begin{equation}
\sum_{g\in\mathcal G_i}P_{g,t}+\sum_{r\in\mathcal R_i}R_{r,t}
-d_{i,t}-\delta_{ib}p_t^{\mathrm{DC}}
=\sum_{\ell\in\mathcal L}A_{i\ell}f_{\ell,t}.
\label{eq:nodal_balance}
\end{equation}
Here \(\delta_{ib}=1\) if \(i=b\), and zero otherwise. This balance holds for all \(i\in\mathcal N\) and \(t\in\mathcal T\). The incidence matrix \(\mathbf A\) has entries \(+1\) and \(-1\) at the
sending and receiving ends, \(o(\ell)\) and \(e(\ell)\). For every hour,
\begin{gather}
f_{\ell,t}=\frac{S_{\mathrm{base}}}{x_\ell}
\bigl(\theta_{o(\ell),t}-\theta_{e(\ell),t}\bigr),
\qquad\ell\in\mathcal L,
\label{eq:base_flow}\\
-F_\ell\leq f_{\ell,t}\leq F_\ell,
\qquad\ell\in\mathcal L,
\label{eq:line_limits}\\
\underline P_g\leq P_{g,t}\leq\overline P_g,
\qquad g\in\mathcal G,
\label{eq:generation_bounds}\\
0\leq R_{r,t}\leq\overline R_{r,t},
\qquad r\in\mathcal R,
\label{eq:profile_bounds}\\
-\pi\leq\theta_{i,t}\leq\pi,
\qquad i\in\mathcal N,
\label{eq:angle_bounds}\\
\theta_{i_{\mathrm{ref}},t}=0.
\label{eq:reference_angle}
\end{gather}
Here \(x_\ell\) and \(F_\ell\) are branch reactance and rating,
\(S_{\mathrm{base}}\) is base power, and \(\overline R_{r,t}\) is hourly
available output. These constraints define intact-network operation.

For the selected non-islanding outages \(c\in\mathcal C\), line outage
distribution factors (LODFs) are computed from the intact network.
Remove the reference-bus row of \(\mathbf A\) to obtain
\(\widetilde{\mathbf A}\), with column \(\widetilde{\mathbf a}_c\).
Define
\begin{gather}
\mathbf D=\operatorname{diag}_{\ell\in\mathcal L}
\left(\frac{S_{\mathrm{base}}}{x_\ell}\right),
\label{eq:branch_susceptance}\\
\mathbf B_{\mathrm{bus}}=\widetilde{\mathbf A}\mathbf D
\widetilde{\mathbf A}^{\mathsf T},
\label{eq:reduced_susceptance}\\
\mathbf h^{(c)}=\mathbf D\widetilde{\mathbf A}^{\mathsf T}
\mathbf B_{\mathrm{bus}}^{-1}\widetilde{\mathbf a}_c,
\label{eq:branch_transfer}\\
L_{\ell c}=
\begin{cases}
h_\ell^{(c)}/(1-h_c^{(c)}),&\ell\ne c,\\
-1,&\ell=c.
\end{cases}
\label{eq:lodf}
\end{gather}
The vector \(\mathbf h^{(c)}\) gives branch-flow responses to a unit
transfer from \(o(c)\) to \(e(c)\). The N$-$1 constraints are
\begin{gather}
f_{\ell,t}^{(c)}=f_{\ell,t}+L_{\ell c}f_{c,t},
\qquad\ell\in\mathcal L,
\label{eq:post_flow}\\
-F_\ell\leq f_{\ell,t}^{(c)}\leq F_\ell,
\qquad\ell\in\mathcal L\setminus\{c\},
\label{eq:preventive_security}
\end{gather}
for every \(c\in\mathcal C\) and \(t\in\mathcal T\); \(L_{cc}=-1\)
sets the outaged line flow to zero. Each outage is evaluated independently
using the same hourly generation dispatch and data center schedule,
representing preventive operation.

\subsection{Hosting Capacity assessment framework}
Let \(z\) collect the operating decisions and \(\mathcal F\) denote the feasible capacities and schedules under these prescribed settings. The objective function of this framework is
\begin{equation}
H^\star=\max_{H\geq0,\,z}\{H\mid(H,z)\in\mathcal F\}.
\label{eq:hosting_capacity}
\end{equation}
The capacity gain from workload flexibility is obtained by comparing two
separate solves at the same connection bus and under the same security
condition:
\begin{equation}
V=H^{\star,\mathrm{Flex}}-H^{\star,\mathrm{Non\text{-}Flex}}.
\label{eq:flexibility_value}
\end{equation}
The N$-$1 solution starts from the normal network operation. After each solve, all selected outages are checked in every hour, and the
two-sided constraint on the branch is added for each violating outage-hour pair.

\section{Case Study and Results Analysis}
\label{sec:case_results}

\subsection{Case Study Setup}
The optimization model was implemented in Julia 1.12.5 and solved with Gurobi 13.0.0 \cite{Gurobi2026}. The study is assessed on a modified IEEE 118-bus system with 186 branches and 118 buses\cite{Pena2018NREL118}. A contingency-prone area is defined to encompass buses 82–96 and 100–112. Buses 92 and 77 are selected as representative data center connection locations within and outside this area, respectively, to investigate the influence of spatial proximity to contingencies. At each hourly time step, N-1 security constraints are imposed for all 38 non-islanding branch outages within the contingency area, ensuring that all remaining in-service branches operate within their thermal limits. The workload profile is constructed from the IN2P3 computing workload trace \cite{CochardSimon2026IN2P3}. A 1,464-hour period covering November and December 2024 is selected for analysis. The hourly batch workload is assumed and calibrated to 20\% of the total arriving workload. Table~\ref{tab:dc_parameters} gives the parameters.
\begin{table}[H]
\centering
\caption{Data Center Configuration}
\label{tab:dc_parameters}
\footnotesize
\begin{tabular}{@{}l@{\hspace{2em}}l@{}}
\toprule
Parameter & Value \\
\midrule
\(\eta\) & \(1.20\) \\
\(p^{\mathrm{idle}}\) & \(1.3~(\mathrm{kW})\) \\
\(p^{\mathrm{peak}}\) & \(1.9~(\mathrm{kW})\) \\
\(\mu\) & \(500~(\text{workload units}/\mathrm{h})\) \\
\(\bar u\) & \(0.80\) \\
\(K\) & \(0,1,2,4,6~(\mathrm{h})\) \\
\bottomrule
\end{tabular}
\end{table}

Two scenarios are established for this case study, considering normal operation and security-constrained operation:
\begin{itemize}
    \item \textbf{Scenario 1 (Non-Flex):} No workload flexibility is allowed.
    \item \textbf{Scenario 2 (Flex):} The batch workload may be deferred within their deadlines window. 
\end{itemize}

\subsection{Impact of Security Constraints and Workload Flexibility}

With workload flexibility, the two sites have nearly identical hosting capacities in the normal operation, differing by only 13.5~MW (Table~\ref{tab:hc_sites}). Enforcing the N$-$1 security constraints widens this gap to 874.5~MW. At bus~92, within the contingency area, hosting capacity decreases by 1415.7~MW, compared with 554.7~MW at bus~77, outside it. With identical condition, the larger reduction at bus~92 demonstrates its greater sensitivity to the transmission line outages. The reduction at bus~77 also shows that the influence of outages extends beyond the contingency area boundary.  

\begin{table}[H]
\caption{Hosting capacity under different scenarios}
\label{tab:hc_sites}
\centering
\footnotesize
\setlength{\tabcolsep}{1.8pt}
\begin{tabular}{@{}lccccc@{}}
\toprule
Bus & Security constraint & Non-Flex (MW) & Flex (MW) & Gain (MW) & Gain (\%)\\
\midrule
92 & - & 5087.9 & 6152.1 & 1064.3 & 20.9\\
   & N$-$1 & 3804.3 & 4736.5 & 932.2 & 24.5\\
\addlinespace[2pt]
77 & - & 5299.2 & 6165.6 & 866.4 & 16.3\\
   & N$-$1 & 4739.0 & 5610.9 & 871.9 & 18.4\\
\bottomrule
\end{tabular}
\end{table}

Nevertheless, workload flexibility provides additional hosting capacity at both sites under N$-$1 constraints: Temporal flexibility adds 932.2~MW at bus~92 and 871.9~MW at bus~77, corresponding to increases of 24.5\% and 18.4\% over inflexibility case. This distinction separates the value of workload scheduling from the suitability of a connection location. At bus~92, flexibility releases considerable additional capacity where the line outages impose the larger capacity reduction; at bus~77, it further increases the capacity available at the less affected site. Workload flexibility improves the use of the network at both locations, while the remaining difference in secure hosting capacity preserves the importance of siting.

\subsection{N-1 Contingencies and Workload Shifting analysis}
\label{sec:critical_analysis}

Screening under non-flexibility identifies November~7 at 20:00 as a
limiting N$-$1 hour at bus 77 and 92. Fig.~\ref{fig:critical_hours}(a)--(b)
compares power consumption with and without workload flexibility: 4736.5~MW at bus~92 and 5610.9~MW at bus~77. At the limiting hour, batch workload deferral reduces power from 3975.8 to 3193.3~MW at bus~92 and from 4709.8 to 3977.9~MW at bus~77. The reductions of 782.5 and
731.9~MW allow the larger DC hosting capacity to match the critical-hour power requirements of the smaller DC Capacity operating without workload flexibility. Thus, additional hosting capacity is obtained by reshaping the demand associated with more computing obligation while retaining the same
critical-hour import.

\begingroup
\setlength{\intextsep}{3pt}
\begin{figure}[H]
\centering
\includegraphics[width=.94\columnwidth]{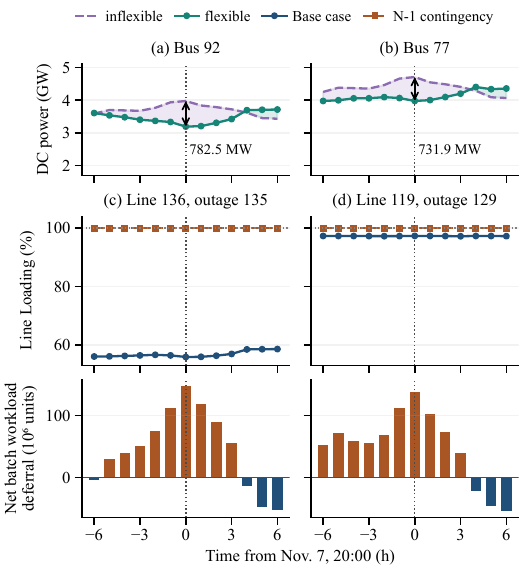}
\caption{Data Center Operation and Transmission Constraints during critical hours.
(a), (b): DC power consumption at bus 77/92 under two scenarios.
(c), (d): line loading and batch workload at two critical lines in normal  and N-1 Contingency operation(positive and negative values indicate net deferral and recovery).}
\label{fig:critical_hours}
\end{figure}
\endgroup

The network constraints underlying these power limits differ between sites.
At bus~92, line~136 (85--89) carries 335.6~MW on the intact network,
only 56\% of its 600~MW rating. An outage of line~135 adds
264.4~MW and exhausts this headroom. At bus~77, line~119 (69--77)
already carries 583.9~MW and the 16.1~MW increase following an outage of
line~129 is sufficient to reach its rating. Thus, the constraint at bus~92 becomes binding through substantial contingency-induced redistribution, whereas that at bus~77 results from a small redistribution onto an already heavily loaded line. Accommodating additional data center demand therefore requires an hourly operating schedule that respects these post-contingency limits..

The bottom panels connect the power reduction to workload scheduling.
At 20:00, net deferral reaches 146.5 million workload units at bus~92
and 137.1 million at bus~77. From midnight through 02:00, execution
exceeds new arrivals and scheduled power rises above the non-flexibility
benchmark as earlier work is recovered. Workload recovery therefore takes place within the changing hourly balance of generation, native demand, and data center power. The flexibility value benefits from the ability to accommodate deferred workload across the scheduling window as well as to reduce demand at the limiting hour.

\makeatletter
\let\savedstartsection\@startsection
\def\@startsection#1#2#3#4#5#6{\savedstartsection{#1}{#2}{#3}{1pt}{1pt}{#6}}
\subsection{Sensitivity to the Workload Delay Window}
\begin{figure}[H]
\centering
\includegraphics[width=.94\columnwidth]{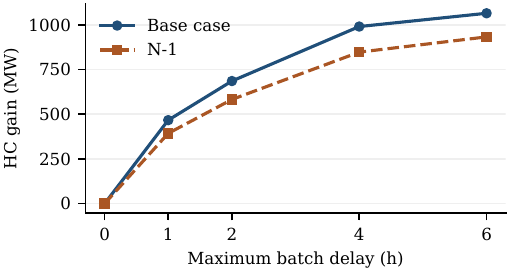}
\caption{Hosting capacity gain at bus~92 as a function of the
maximum permitted batch workload delay.}
\label{fig:delay_window}
\end{figure}

\let\@startsection\savedstartsection
\makeatother
Although delay-related costs are not quantified in this study, their impact on service performance and economic value is an important consideration when selecting a practical workload delay window. The sensitivity to the delay window shows that at bus~92, allowing one hour of batch workload deferral increases hosting capacity by 393.2~MW under N$-$1 constraints, capturing 42.2\% of the gain available with a six-hour delay window (Fig.~\ref{fig:delay_window}). Extending the delay window from two to four hours adds 264.7~MW, whereas extending it from four to six hours adds only 85.1~MW. The corresponding marginal benefit falls from 132.4 to 42.6~MW per additional hour of permitted delay window. A four-hour deadline therefore captures 90.9\% of the modelled six-hour N$-$1 gain, indicating that most of the capacity benefit can be obtained with a shorter delay allowance. The base case exhibits a similar pattern, with gains of 304.6 and 74.3~MW over the same two intervals. These results show diminishing capacity benefits from extending the batch delay window under  both normal and security conditions. 

\section{Conclusion}
This paper develops a data center hosting capacity framework that links
data center connection capacity to its computing obligation and hourly
power demand under N$-$1 security constraints. The case study shows that workload flexibility increases secure hosting capacity at data center sites, while its incremental value and the achievable connection size remain distinct: bus~92 obtains the larger gain from workload temporal flexibility. Connection planning therefore needs to evaluate the capacity enabled by scheduling alongside the data center connection bus's location relative to the contingency-prone area.

The operational analysis shows that additional capacity depends on both
reducing demand at limiting hours and accommodating the deferred work
within its deadline. The relevant flexibility commitment is thus a
feasible execution schedule over the delay window, rather than an isolated
power reduction. At bus~92, four hours of deferral capture 90.9\% of the
six-hour N$-$1 gain, indicating diminishing returns from longer deadlines in
this case. Incorporating delay-related service costs and uncertain
workload arrivals would extend this capacity assessment to the selection
of practical scheduling commitments.

\bibliographystyle{IEEEtran}
\bibliography{references}

\end{document}